\documentclass[aps,amsmath,amssymb]{revtex4}
\usepackage{graphicx}

\begin{document}


\title{\Large \bf Vector-pseudoscalar two-meson distribution
amplitudes in three-body $B$ meson decays
}
\date{\today}
\author{\large \bf  Chuan-Hung Chen$^{a}$\footnote{Email:
phychen@mail.ncku.edu.tw} and Hsiang-nan
Li$^{a,b}$\footnote{Email: hnli@phys.sinica.edu.tw}}

\vskip1.0cm

\affiliation{$^{a}$Department of Physics, National Cheng-Kung University,
Tainan, Taiwan 701, Republic of China}
\affiliation{$^b$Institute of Physics, Academia Sinica,
Taipei, Taiwan 115, Republic of China}

\begin{abstract}

We study three-body nonleptonic decays $B\to VVP$ by introducing
two-meson distribution amplitudes for the vector-pseudoscalar
pair, such that the analysis is simplified into the one for
two-body decays. The twist-2 and twist-3 $\phi K$ two-meson
distribution amplitudes, associated with longitudinally and
transversely polarized $\phi$ mesons, are constrained by the
experimental data of the $\tau\to\phi K\nu$ and $B\to\phi K\gamma$
branching ratios. We then predict the $B\to\phi K\gamma$ and
$B\to\phi\phi K$ decay spectra in the $\phi K$ invariant mass.
Since the resonant contribution in the $\phi K$ channel is
negligible, the above decay spectra provide a clean test for the
application of two-meson distribution amplitudes to three-body $B$
meson decays.

\end{abstract}
\maketitle
\vskip 1.0cm

Viewing the experimental progress on three-body nonleptonic $B$
meson decays \cite{Belle,Bar}, it is urgent to construct a
corresponding framework. In \cite{CL02} we have proposed a
formalism based on the collinear factorization theorem in
perturbative QCD (PQCD), in which new nonperturbative inputs, the
two-meson distribution amplitudes, were introduced \cite{MP}. On
one hand, a direct evaluation of hard kernels for three-body
decays, which contain two virtual gluons at lowest order, is not
practical due to the enormous number of diagrams. On the other
hand, the region with the two gluons being hard simultaneously is
power-suppressed and not important. Therefore, the new
nonperturbative inputs are necessary for catching dominant
contributions in a simple manner. In our formalism the collinear
factorization formula for a $B\to h_1h_2h_3$ decay amplitude is
written, in general, as
\begin{eqnarray}
{\cal M}=\Phi_B\otimes H\otimes \Phi_{h_1h_2}\otimes\Phi_{h_3}\;,
\label{b3d}
\end{eqnarray}
where $\Phi_{B,h_3}$ are the $B,h_3$ meson distribution
amplitudes, $\Phi_{h_1h_2}$ the $h_1h_2$ two-meson distribution
amplitude, and $\otimes$ represents the convolution in
longitudinal momentum fractions $x$. $\Phi_{h_1h_2}$ and
$\Phi_{h_3}$ include not only the twist-2 (leading-twist), but
two-parton twist-3 (next-to-leading-twist) components. The
computation of the hard kernel $H$, basically the same as in
two-body $B$ meson decays, is restricted to leading order in the
coupling constant $\alpha_s$ so far.

There are two types of factorization theorems: collinear
factorization \cite{BL,BFL,MR,DM,CZS} and $k_T$ factorization
\cite{BS,NL}. For a comparison of the two types of theorems, refer
to \cite{Keum,Li03}. Collinear factorization works, if it does not
develop an end-point singularity from $x\to 0$. If it does,
collinear factorization breaks down, and $k_T$ factorization is
more appropriate. It has been known that collinear factorization
of charmed and charmless two-body $B$ meson decays suffers the
end-point singularities \cite{BBNS}. This is the motivation to
develop the PQCD formalism for two-body $B$ meson decays based on
$k_T$ factorization \cite{YL,KLS,LUY}. This approach has been
shown to be infrared-finite, gauge-invariant, and consistent with
the factorization assumption in the heavy-quark limit
\cite{CLY,LY1,LU}. For three-body $B$ meson decays, the end-point
singularities are smeared by the two-meson invariant mass
\cite{CL02}, and collinear factorization in Eq.~(\ref{b3d}) holds.
Moreover, it has been demonstrated that both nonresonant
contributions and resonant contributions through two-body channels
can be included by means of an appropriate parametrization of
$\Phi_{h_1h_2}$ \cite{CL02}.

One of the challenges in the studies of three-body heavy meson
decays is the evaluation of the matrix elements for heavy meson
transition into two hadrons. There are already several theoretical
approaches to this subject in the literature. The naive
factorization \cite{BSW} for three-body $B$ meson decays has been
adopted in \cite{CHST}, in which the $B$ meson transition into two
hadrons was simply parameterized by a power-law behavior and then
fit to experimental data. The matrix elements for the above
transition were calculated using the pole model
\cite{BFOPP,cheng,LGT}, in which intermediate-state decays into
two hadrons were described by effective weak and strong
Lagrangians. The naive factorization has been
improved in a so-called QCD-factorization framework \cite{W}.
However, only the current-produced amplitudes, i.e., those which
can be expressed as products of two form factors in the
factorization limit, were studied. The challenging subject of the
$B$ meson transition into two hadrons was not addressed \cite{W}.
Compared to the above methods, our approach does not rely on the
naive factorization, since the nonfactorizable contribution is
taken into account through nonfactorizable hard kernels. It is 
complete in the sense that various topologies of amplitudes, such 
as the $B$ meson transition into two hadrons and the 
current-induced one, are analyzed in the same framework. It is also 
more systematic, because sub-leading corrections can be evaluated 
order by order in $\alpha_s$ and power by power in the ratios 
$w/m_B$ and $m_{h_3}/m_B$, where $w$ is the invariant mass of the 
two-meson system, and $m_B$ ($m_{h_3}$) the $B$ ($h_3$) meson mass.

In \cite{CL02} we have applied Eq.~(\ref{b3d}) to the modes, in
which both $h_1$ and $h_2$ are pseudoscalar mesons $P$. The modes
with $h_1$ being a vector meson $V$ and $h_2$ a pseudoscalar meson
$P$ have been observed recently \cite{Bel03}. Hence, we shall
extend our formalism to three-body decays involving the $B\to VP$
transition, taking $B\to\phi\phi K$ as an example. We
shall first define the $\phi K$ two-meson distribution amplitudes,
which are more complicated than the $PP$ ones. A simple
parametrization is then proposed, and constrained by the
experimental data of the $\tau\to\phi K\nu$ and $B\to\phi K\gamma$
branching ratios. Afterwards, we predict the decay spectra of the
$B\to\phi K\gamma$ and $B\to\phi\phi K$ modes in the $\phi K$
invariant mass. The resonant contribution through the $\phi K$
channel is expected to be negligible: the $K_1(1650)$,
$K_2(1770)$, and $K(1830)$ mesons decay into the $\phi K$ pair
with the branching ratios not yet available in \cite{PDG}.
Therefore, the above spectra provide a clean test for the
application of two-meson distribution amplitudes to three-body $B$
meson decays.

Label the momenta of the $\phi$ and $K$ mesons from the $B$ meson
transition as $P_1$ and $P_2$, respectively. The $B$ meson
momentum $P_B$ and the total momentum of the $\phi K$ pair,
$P=P_1+P_2$, are chosen, in the light-cone coordinates, as
\begin{eqnarray}
P_B=\frac{m_B}{\sqrt{2}}(1,1,{\bf 0}_T)\;,\;\;\;\;
P=\frac{m_B}{\sqrt{2}}(1,\eta,{\bf 0}_T)\;,
\end{eqnarray}
with the variable $\eta=w^2/m_B^2$. Define $\zeta=P_1^+/P^+$ as
the $\phi$ meson momentum fraction and $r_\phi=m_\phi/m_B$ as the
$\phi$ meson-$B$ meson mass ratio, in terms of which the other
kinematic variables are expressed as
\begin{eqnarray}
& &P_2^+=(1-\zeta)P^+\;,\;\;\;\; P_1^-=[(1-\zeta)\eta+r_\phi^2]
P^+\;,\;\;\;\; P_2^-=(\zeta\eta-r_\phi^2) P^+\;,\nonumber\\
& &P_{1}^x=-P_{2}^x=\sqrt{(\zeta w^2-m_\phi^2)(1-\zeta)}\;,
\;\;\;\;(P_1^x)^2=(P_2^x)^2\equiv P_T^2\;.
\label{ptm}
\end{eqnarray}
The polarization vectors $\epsilon(\phi)$ of the $\phi$ meson are
obtained from the orthogonality $\epsilon(\phi)\cdot P_1=0$ and from
the normalization $\epsilon(\phi)^2=-1$. The exact expressions are
given, in the light-cone coordinates
$\epsilon=(\epsilon^+,\epsilon^-,\epsilon^x,\epsilon^y)$, by
\begin{eqnarray}
\epsilon_{L}(\phi)&=&\frac{1}{r_\phi}\left(
\frac{\zeta[\zeta+(1-\zeta)\eta+r_\phi^2]-2r_\phi^2}
{\sqrt{2}\sqrt{[\zeta+(1-\zeta)\eta+r_\phi^2]^2-4r_\phi^2}},
\frac{[(1-\zeta)\eta+r_\phi^2][\zeta+(1-\zeta)\eta+r_\phi^2]-2r_\phi^2}
{\sqrt{2}\sqrt{[\zeta+(1-\zeta)\eta+r_\phi^2]^2-4r_\phi^2}},\right.
\nonumber\\
& &\left.
\frac{[\zeta+(1-\zeta)\eta+r_\phi^2]\sqrt{(\zeta\eta-r_\phi^2)(1-\zeta)}}
{\sqrt{[\zeta+(1-\zeta)\eta+r_\phi^2]^2-4r_\phi^2}}, 0\right)\;,
\nonumber\\
\epsilon_{T}^{(1)}(\phi)&=&\left(
-\frac{\sqrt{2}\sqrt{(\zeta\eta-r_\phi^2)(1-\zeta)}}
{\sqrt{[\zeta+(1-\zeta)\eta+r_\phi^2]^2-4r_\phi^2}},
\frac{\sqrt{2}\sqrt{(\zeta\eta-r_\phi^2)(1-\zeta)}}
{\sqrt{[\zeta+(1-\zeta)\eta+r_\phi^2]^2-4r_\phi^2}},\right.
\nonumber\\
& &\left.
\frac{[\zeta-(1-\zeta)\eta-r_\phi^2]}
{\sqrt{[\zeta+(1-\zeta)\eta+r_\phi^2]^2-4r_\phi^2}}, 0\right)\;,
\nonumber\\
\epsilon_{T}^{(2)}(\phi)&=&(0,0,0,1)\;. \label{tp2}
\end{eqnarray}
The terms proportional to $r_\phi$ will be neglected eventually.
The kaon is treated as a massless particle. The $\phi$ meson
emitted from the weak vertex then carries the momentum
$P_3=(m_B/\sqrt{2})(0,1-\eta,{\bf 0}_T)$. Another equivalent, but
more general, representation of $\epsilon(\phi)$ is given by
\begin{eqnarray}
\epsilon^{\pm}_{L}(\phi)&=&\frac{P^{2}_{T}\pm
P^{\pm}_{1}(P^{+}_{1}-P^{-}_{1})}{\sqrt{2}m_{\phi}p}\;,\;\;\;\;
    \epsilon_{L}^x(\phi)=\frac{P_{T}(P^{+}_{1}+P^{-}_{1})}
{\sqrt{2}m_{\phi}p}\;,
    \nonumber \\
            \epsilon^{(1)\pm}_{T}(\phi)&=& \frac{\mp
            P_{T}}{\sqrt{2}p}\;,\;\;\;\;
            \ \ \  \epsilon^{(1)x}_{T}(\phi)=
            \frac{P^{+}_{1}-P^{-}_{1}}{\sqrt{2}p},
\end{eqnarray}
with $p=\sqrt{P^{2}_{T}+(P^{+}_{1}-P^{-}_{1})^2/2}$.

The three-body $B$ meson decays are dominated by the contribution
from the region, in which the $\phi K$ pair possesses the
invariant mass $w^2\sim O(\bar\Lambda m_B)$ \cite{CL02},
$\bar\Lambda$ representing a hadronic scale. The orders of
magnitude of the components,
\begin{eqnarray}
P^+\sim O(m_B)\;,\;\;\;\; P^-\sim O(\bar\Lambda)\;,\;\;\;\;
P_T\sim O(\sqrt{\bar\Lambda m_B})\;, \label{p2m}
\end{eqnarray}
are then implied. It is easy to obtain the power counting rules of
the polarization vectors from Eq.~(\ref{tp2}),
\begin{eqnarray}
& &\epsilon_{L}^+(\phi)\sim \frac{1}{r_\phi}O(1)\;,\;\;\;\;
\epsilon_{L}^x(\phi)\sim
\frac{1}{r_\phi}O\left(\sqrt{\bar\Lambda/m_B}\right)\;,\;\;\;\;
\epsilon_{L}^-(\phi)\sim \frac{1}{r_\phi}O(\bar\Lambda/m_B)\;,
\nonumber\\
& &\epsilon_{T}^{(1)+}(\phi)\sim \epsilon_{T}^{(1)-}(\phi)\sim
O\left(\sqrt{\bar\Lambda/m_B}\right)\;,\;\;\;\;
\epsilon_{T}^{(1)x}(\phi)\sim O(1)\;. \label{ptp1}
\end{eqnarray}
In the heavy-quark limit the hierarchy $P^+\gg P_{T}\gg P^-$
corresponds to a collinear configuration, and suggests the
employment of the new nonperturbative inputs, the $\phi K$
two-meson distribution amplitudes. For the $PP$ system, there is
only a single twist-2 distribution amplitude associated with the
structure $\gamma_\mu$, and two two-parton twist-3 distribution
amplitudes associated with the structures $I$ (the identity) and
$\sigma_{\mu\nu}$ \cite{CL02,MP,M}. Here a higher-twist
distribution amplitude means that its contribution is suppressed
by powers of $w/m_B$. For the $VP$ system, the relevant structures
are more complicated: three twist-2 distribution amplitudes are
associated with $\gamma_\mu\gamma_5$ and
$\sigma_{\mu\nu}\gamma_5$, and five twist-3 distribution amplitudes
with $\gamma_\mu\gamma_5$, $\sigma_{\mu\nu}\gamma_5$, $\gamma_5$ 
and $\gamma_\mu$. To decompose the two-meson
distribution amplitudes into the components of different twists,
we introduce the polarization vectors of the $\phi K$ system,
\begin{eqnarray}
\epsilon_{L}=\frac{1}{\sqrt{2\eta}}(1,-\eta,0,0)\;,\;\;\;\;
\epsilon_{T}^{(1)}=(0,0,1,0)\;,\;\;\;\;\epsilon_{T}^{(2)}=(0,0,0,1)\;.
\label{pv}
\end{eqnarray}

A two-pion distribution amplitude has been related to the pion
distribution amplitude through a perturbative calculation of the
process $\gamma\gamma^*\to \pi^+\pi^-$ at large invariant mass
$w^2$ \cite{DFK}. In this work we adopt a similar trick: we
calculate perturbatively the matrix elements,
\begin{eqnarray}
\langle \phi(P_1,\epsilon(\phi))K^+(P_2)|\bar u(y^-)\Gamma
s(0)|0\rangle\;,
\end{eqnarray}
using the $\phi$ meson and kaon distribution
amplitudes up to twist 3 \cite{PB1,PB2}, where $\Gamma$ represents
a structure among $I$, $\gamma_5$, $\gamma_\mu$,
$\gamma_\mu\gamma_5$ and $\sigma_{\mu\nu}\gamma_5$. The matrix
elements can be expressed as the products of the corresponding
form factors with the kinematic factors. For example, the matrix
element for $\Gamma=\gamma_\mu\gamma_5$ is written as the product
of the form factor $F_\parallel$ with the kinematic factor
$(P_1-P_2)_\mu$. The kinematic factors are then approximated in terms
of the momentum $P$ and the polarization vectors $\epsilon$ of the
$\phi K$ system according to the power counting rules in
Eqs.~(\ref{p2m}) and (\ref{ptp1}). The resultant $\zeta$-dependent
coefficients in the approximation contribute to the $\zeta$ dependence
of the $\phi K$ two-meson distribution amplitudes.

We then derive the decomposition up to $O(w/m_B)$,
\begin{eqnarray}
\langle \phi K^+|\bar u(y^-)\gamma_\mu\gamma_5
s(0)|0\rangle&=&P_{\mu} \int_0^1 dz e^{iz P\cdot y}
\Phi_\parallel(z,\zeta,w)\;,
\label{pa}\\
\langle \phi K^+|\bar u(y^-)\sigma_{\mu\nu}\gamma_5
s(0)|0\rangle&=&-i\left\{(\epsilon_{T\mu}P_{\nu}-\epsilon_{T\nu}P_{\mu})
\int_0^1 dz e^{iz P\cdot
y}\Phi_T(z,\zeta,w)\right.\nonumber\\
& &\left.+\frac{2}{w}(P_{1\mu}P_{2\nu}-P_{1\nu}P_{2\mu})\int_0^1
dz e^{iz P\cdot y}\Phi_3(z,\zeta,w)\right\}\;,\label{pt}\\
\langle\phi K^+|\bar u(y^-)\gamma_5 s(0)|0\rangle&=& w\int_0^1 dz
e^{iz P\cdot y}\Phi_p(z,\zeta,w)\;,
\label{pap}\\
\langle\phi K^+|\bar u(y^-)\gamma_\mu
s(0)|0\rangle&=& i\frac{w}{P\cdot n_-}\epsilon_{\mu\nu\rho\sigma}
\epsilon_{T}^{\nu}P^{\rho}n_-^{\sigma}\int_0^1 dz e^{iz P\cdot
y}\Phi_v(z,\zeta,w)\;, \label{paa}\\
\langle\phi K^+|\bar u(y^-)I
s(0)|0\rangle&=&0\;,
\end{eqnarray}
where $z$ is the momentum fraction carried by the spectator $u$
quark, and $n_-=(0,1,{\bf 0}_T)$ a null vector. We have adopted
the convention $\epsilon^{0123}=1$ for the Levi-Civita tensor
$\epsilon^{\mu\nu\rho\delta}$. The above decomposition applies to
other $VP$ systems, such as $K^*\pi$, $\rho K$, $\cdots$.

Below we present some details of the expansion of the kinematic
factors. For Eq.~(\ref{pa}), we have applied
\begin{eqnarray}
(P_1-P_2)_\mu\approx(2\zeta-1)P_\mu\;,
\label{p1p2}
\end{eqnarray}
where the coefficient $2\zeta-1$ is absorbed into the distribution
amplitude $\Phi_\parallel$, giving its $\zeta$ dependence.
Similarly, we have approximated the
kinematic factor for the matrix element in Eq.~(\ref{pt}),
\begin{eqnarray}
\epsilon_{T\mu}(\phi)P_{1\nu}-\epsilon_{T\nu}(\phi)P_{1\mu}
\approx\zeta (\epsilon_{T\mu}P_{\nu}-\epsilon_{T\nu}P_{\mu})\;,
\label{tptp}
\end{eqnarray}
where the coefficient $\zeta$ is absorbed into $\Phi_T$, and
$\epsilon_{T\mu}$ is a transverse polarization vector
of the $\phi K$ system. The contribution from another
distribution amplitude $\Phi_3$ can be
combined with that from $\Phi_T$ via the approximation,
\begin{eqnarray}
\frac{2}{w}\left(P_{1\mu}P_{2\nu}-P_{1\nu}P_{2\mu}\right)\approx
2\sqrt{\zeta(1-\zeta)}(\epsilon^{(1)}_{T\mu}P_{\nu}
-\epsilon^{(1)}_{T\nu}P_{\mu})\;, \label{epep}
\end{eqnarray}
where the coefficient $\sqrt{\zeta(1-\zeta)}$ comes from $P_1^x$
in the $m_\phi\to 0$ limit. Since the branching ratio is a sum
over the transverse polarizations $\epsilon^{(1)}_{T\mu}$ and
$\epsilon^{(2)}_{T\mu}$, we omit the coefficient 2, and replace
$\epsilon^{(1)}_{T\mu}$ by the two possible $\epsilon_{T\mu}$. We
have employed the approximation for the matrix element in
Eq.~(\ref{paa}),
\begin{eqnarray}
\frac{2}{w}\epsilon_{\mu\nu\rho\sigma}
\epsilon_{T}^{\nu}(\phi)P_1^{\rho}P_2^{\sigma}\approx
\frac{w}{P\cdot n_-}\zeta\epsilon_{\mu\nu\rho\sigma}
\epsilon_{T}^{\nu}P^{\rho}n_-^{\sigma}\;.
\end{eqnarray}
For this structure, the $\phi$ meson emitted from the weak vertex 
must carry a transverse polarization, and a
non-vanishing hard kernel demands that the subscript $\mu$ denotes
a transverse component. The coefficient $\zeta$ is then the sum of
$\zeta$, $\zeta-1$, and $1-\zeta$ from the combinations
($\epsilon_{T}^{(1)\nu}(\phi)=\epsilon_{T}^{(1)\perp}(\phi)$,
$P^\rho=P^+$, $P_2^\sigma=P_2^-$),
($\epsilon_{T}^{(1)\nu}(\phi)=\epsilon_{T}^{(1)\perp}(\phi)$,
$P^\rho=P^-$, $P_2^\sigma=P_2^+$), and
($\epsilon_{T}^{(1)\nu}(\phi)=\epsilon_{T}^{(1)-}(\phi)$,
$P^\rho=P^+$, $P_2^\sigma=P_2^\perp$), respectively. A coefficient
2 for the last combination has been omitted for the same reason.

Our strategy does not provide the $z$ dependence.
Assuming the $z$ dependence of each $\Phi_i(z,\zeta,w)$ to be
asymptotic,
we propose the parametrization,
\begin{eqnarray}
\Phi_\parallel(z,\zeta,w)&=&\frac{3F_\parallel(w)}{\sqrt{2N_c}}
z(1-z)(2\zeta-1)\;, \;\;\;\;
\nonumber\\
\Phi_T(z,\zeta,w)&=&\frac{3F_T(w)}{\sqrt{2N_c}}z(1-z)\zeta\;,\;\;\;\;
\nonumber\\
\Phi_3(z,\zeta,w)&=&\frac{3F_3(w)}{\sqrt{2N_c}}z(1-z) \;,
\nonumber\\
\Phi_p(z,\zeta,w)&=&\frac{3F_p(w)}{\sqrt{2N_c}}z(1-z)\;,
\nonumber\\
\Phi_v(z,\zeta,w)&=&\frac{3F_v(w)}{\sqrt{2N_c}}z(1-z)\zeta\;.
\label{2pi}
\end{eqnarray}
The time-like form factors $F_{\parallel,T,3,p,v}(w)$ define the
normalization of the $\phi K$ two-meson distribution amplitudes.
Note that these form factors are normalized to
$F_{\parallel,T,3,p,v}(m_\phi)=1$ in order to respect the
kinematic threshold of decay spectra. Our strategy also reveals
the power behaviors of the form factors in the asymptotic region
with large $w$, $F_{\parallel,T}(w)\sim 1/w^2$ and
$F_{3,p,v}(w)\sim m_0/w^3$, $m_0\approx 1.7$ GeV \cite{KLS,HHZ}
being the chiral scale. Therefore, we further parameterize the
form factors in the whole range of $w$ for the evaluation of the
nonresonant contribution:
\begin{eqnarray}
F_\parallel(w)&=&\frac{m_\parallel^2}{(w-m_\phi)^2+m_\parallel^2}\;,
\nonumber\\
F_T(w)&=&\frac{m_T^2}{(w-m_\phi)^2+m_T^2}\;,\nonumber\\
F_3(w)&=&F_p(w)=\frac{m_0m_\parallel^2}{(w-m_\phi)^3+m_0m_\parallel^2}\;,
\nonumber\\
F_v(w)&=&\frac{m_0m_T^2}{(w-m_\phi)^3+m_0m_T^2}\;, \label{non}
\end{eqnarray}
where the two free parameters $m_{\parallel,T}$, expected to be
few GeV \cite{CL02}, are determined by the fit to the measured
$\tau \to \phi K\nu$ and $B\to \phi K \gamma$ branching ratios
\cite{PDG}. The form factors depending on the parameter
$m_\parallel$ ($m_T$) are associated with the longitudinally
(transversely) polarized $\phi$ meson.

We stress that Eqs.~(\ref{pa}) and (\ref{pt}) contain not only the 
twist-2 distribution amplitudes, but the twist-3 ones. 
The expansion in Eq.~(\ref{p1p2}) corresponding to the component 
$\mu=\perp$ generates
\begin{eqnarray}
& &w\epsilon_{T\mu}\int_0^1 dz e^{iz P\cdot y}
\Phi_a(z,\zeta,w)\;,\nonumber\\
& &\Phi_a(z,\zeta,w)=\frac{3F_\parallel(w^2)}{\sqrt{2N_c}}
z(1-z)\sqrt{\zeta(1-\zeta)}\;.
\label{p1a}
\end{eqnarray}
Similarly, we extract two twist-3 distribution amplitudes from
Eqs.~(\ref{tptp}) and (\ref{epep}) corresponding to the components 
$\mu,\nu=+,-$, given by
\begin{eqnarray}
& &(\epsilon_{L\mu}P_{\nu}-\epsilon_{L\nu}P_{\mu}) \int_0^1 dz e^{iz
P\cdot y}\left[(2\zeta-1)\Phi_3(z,\zeta,w)-\Phi_t(z,\zeta,w)\right]
\;,\nonumber\\
& &\Phi_t(z,\zeta,w)=\frac{3F_T(w^2)}{\sqrt{2N_c}}z(1-z)
\sqrt{\zeta(1-\zeta)}\;.\label{3t}
\end{eqnarray}
For the $\phi K$ system, the above twist-3 distribution amplitudes
lead to smaller contributions compared to $\Phi_p$
and $\Phi_v$, and have been ignored: because of $m_\phi^2/w^2\sim 1$, 
the range in Eq.~(\ref{rzeta}) below indicates $\zeta\sim 1$, and 
that the contribution from $\Phi_a$ is suppressed by the factor 
$\sqrt{1-\zeta}$. There exists a strong cancellation between
$\Phi_3$ and $\Phi_t$ in Eq.~(\ref{3t}).
For other systems, such as $\rho K$, 
these twist-3 distribution amplitudes could be
numerically important due to $m_\rho^2/w^2\ll 1$ in this
case. 

For the $B$ meson distribution amplitude, we use the model \cite{KLS},
\begin{eqnarray}
\Phi_{B}(x)=N_Bx^2(1-x)^2\exp
\left[-\frac{1}{2}\left(\frac{xm_B}{\omega_B}\right)^2\right]\;,
\label{phib}
\end{eqnarray}
with the shape parameter $\omega_{B}=0.40\pm 0.04$ GeV \cite{TLS},
and the normalization constant $N_{B}$ being related to the decay
constant $f_{B}=190$ MeV (in the convention $f_{\pi}=130$ MeV) via
$\int_{0}^{1}\Phi_{B}(x)dx=f_{B}/(2\sqrt{2N_c})$. The range of
$\omega_B$ is determined from a fit to the values of the $B\to\pi$
form factor from light-cone sum rules \cite{KR,PB3}. The above
$\Phi_B$ is identified as $\Phi_+$ of the two leading-twist $B$
meson distribution amplitudes $\Phi_\pm$ defined in \cite{GN,DS}.
Equation (\ref{phib}), vanishing at $x\to 0$, is consistent with
the behavior required by equations of motion \cite{KKQT}. It has
been shown that the $B$ meson distribution amplitude is
normalizable in $k_T$ factorization theorem \cite{LL04}, contrary
to the conclusion drawn in the framework of collinear
factorization theorem \cite{Neu03,BIK}. Another distribution
amplitude ${\bar\Phi}_B$, identified as
${\bar\Phi}_B=(\Phi_{-}-\Phi_{+})/\sqrt{2}$ with a zero
normalization, contributes at the next-to-leading power of
$\bar\Lambda/m_B$ \cite{TLS}. It has been verified numerically
\cite{LMY} that the contribution to the $B\to\pi$ form factor from
$\Phi_B$ is much larger than from ${\bar\Phi}_B$.

In summary, we calculate the hard kernels by contracting the
quark-level diagrams with the matrix elements,
\begin{eqnarray}
\langle 0|{\bar b}(0)_{l}d(y^-)_{j}|B(P_B)\rangle &=&
\frac{1}{\sqrt{2N_c}}\int_0^1 dx e^{-ixP\cdot
y}[(\not{P}_B+m_B)\gamma_5]_{lj}\Phi_B(x)\;,
\nonumber\\
\langle \phi K(P,\epsilon_L)|\bar u(y^-)_js(0)_l|0\rangle &=
&\frac{1}{\sqrt{2N_c}}\int_0^1 dz e^{izP\cdot
y}\left\{(\gamma_5\not P)_{lj}\Phi_\parallel(z,\zeta,w)
+(\gamma_5)_{lj}
w\Phi_p(z,\zeta,w)\right\}\;, \nonumber\\
\langle \phi K(P,\epsilon_T)|\bar u(y^-)_js(0)_l|0\rangle &=
&\frac{1}{\sqrt{2N_c}}\int_0^1 dz e^{izP\cdot y}\left\{
(\gamma_5\not\epsilon_T\not P)_{lj} \left[\Phi_T(z,\zeta,w)
+\Phi_3(z,\zeta,w)\sqrt{\zeta(1-\zeta)}\right]\right.\nonumber\\
& &\left.
+i\frac{w}{P\cdot n_-}\epsilon_{\mu\nu\rho\sigma}
(\gamma^\mu)_{lj}\epsilon_T^\nu P^\rho
n_-^\sigma\Phi_v(z,\zeta,w)\right\}\;, \label{s2}
\end{eqnarray}
which follow Eqs.~(\ref{pa})-(\ref{paa}). The calculation of hard
kernels is as simple as of two-body decays. It is
observed that the distribution amplitudes $\Phi_{\parallel,T,3}$
give leading contributions, and those from $\Phi_{p,v}$ are
suppressed by a power of $w/m_B$.


The $\tau \to \phi K\nu$ differential decay rate in the $\phi K$
invariant mass is written as
\begin{eqnarray}
\frac{d\Gamma}{d w}={G^{2}_{F} m^{4}_{\tau}\over 384 \pi^3}
|V_{us}|^2 \sqrt{\eta}(1-\eta)^2F^{2}_{\parallel}\;,
\label{tau}
\end{eqnarray}
with $m_\tau$ being the $\tau$ lepton mass, and $V_{us}$ the
Cabibbo-Kobayashi-Maskawa matrix element.
The $B\to\phi K\gamma$ decay spectrum is written as
\begin{eqnarray}
\frac{d\Gamma}{d w}=\frac{G_F^2m_B^4}{256\pi^3}\sqrt{\eta}
(1-\eta) \int_{m_\phi^2/w^2}^1d\zeta |{\cal M}(\zeta,w)|^2\;,
\label{dfc}
\end{eqnarray}
with the amplitude,
\begin{eqnarray}
{\cal M}(\zeta,w)&=&\frac{e}{4\pi^2}V_{ts}V_{tb}m_{b}
{\cal A}(\zeta,w)\;,\nonumber\\
{\cal A}(\zeta,w)&\equiv&\langle \phi K| \bar{b} \sigma^{\mu \nu}
\epsilon_{\mu}(\gamma)
q_{\nu} (1-\gamma_{5}) s|B \rangle \nonumber\\
&=&  8\pi C_{F} m^2_{B}
\epsilon_{T}(\gamma)\cdot \epsilon_{T} \int^{1}_{0} dx_{1}
dz\frac{\Phi_{B}(x_{1}) }{x_{1}zm^{2}_{B}+P_{T}^2}
\nonumber \\
&& \times \Bigg\{
\Big[(1+z)\left(\Phi_{T}(z,\zeta,w)
+\Phi_{3}(z,\zeta,w)\sqrt{\zeta(1-\zeta)}\right)
+\sqrt{\eta}(1-2z)\Phi_{v}(z,\zeta,w)
\Big] \frac{\alpha _{s}(
t^{(1)}_{e})C_{7\gamma}^{{\rm eff}}(t^{(1)})}
{zm^2_{B}+P_{T}^2} \nonumber \\
&& +\sqrt{\eta} \Phi_{v}(z,\zeta,w)
\frac{\alpha _{s}( t^{(2)}_{e})C_{7\gamma}^{{\rm eff}}(t^{(2)})}
{x_{1} m^2_{B}
} \Bigg\}\;. \label{gamma}
\end{eqnarray}
$\epsilon(\gamma)$ and $q_\nu$ represent the photon polarization
vectors and the photon momentum, respectively. $m_b$ is the $b$
quark mass, and $C^{\rm eff}_{7\gamma}$ the corresponding
effective Wilson coefficient \cite{BBL}. All the terms of
$O(\eta)$ in ${\cal M}$ have been neglected for consistency.
The requirement $P_T^2\ge 0$
leads to the bounds of $\zeta$ as shown in Eq.~(\ref{dfc}),
\begin{eqnarray}
m_\phi^2/w^2 \le \zeta \le 1\;.\label{rzeta}
\end{eqnarray}
The hard scales are chosen as the maximal virtuality in each
quark-level diagram \cite{CL02,KLS},
\begin{eqnarray}
t^{(1)}=\sqrt{zm_B^2+P_T^2}\;,\;\;\;\; t^{(2)}=\sqrt{x_1m_B^2+
P_T^2}\;.
\end{eqnarray}
The above collinear factorization formula is well-defined, since
the invariant mass of the two-pion system, appearing through
$P_T$, smears the end-point singularities from $z\to 0$. Even if
one adopts a model of the $B$ meson distribution amplitude, which
vanishes only linearly in $x_1$, Eq.~(\ref{gamma}) is still
well-defined due to the presence of $P_T$.

\begin{figure}[htbp]
\includegraphics*[width=3.0
  in]{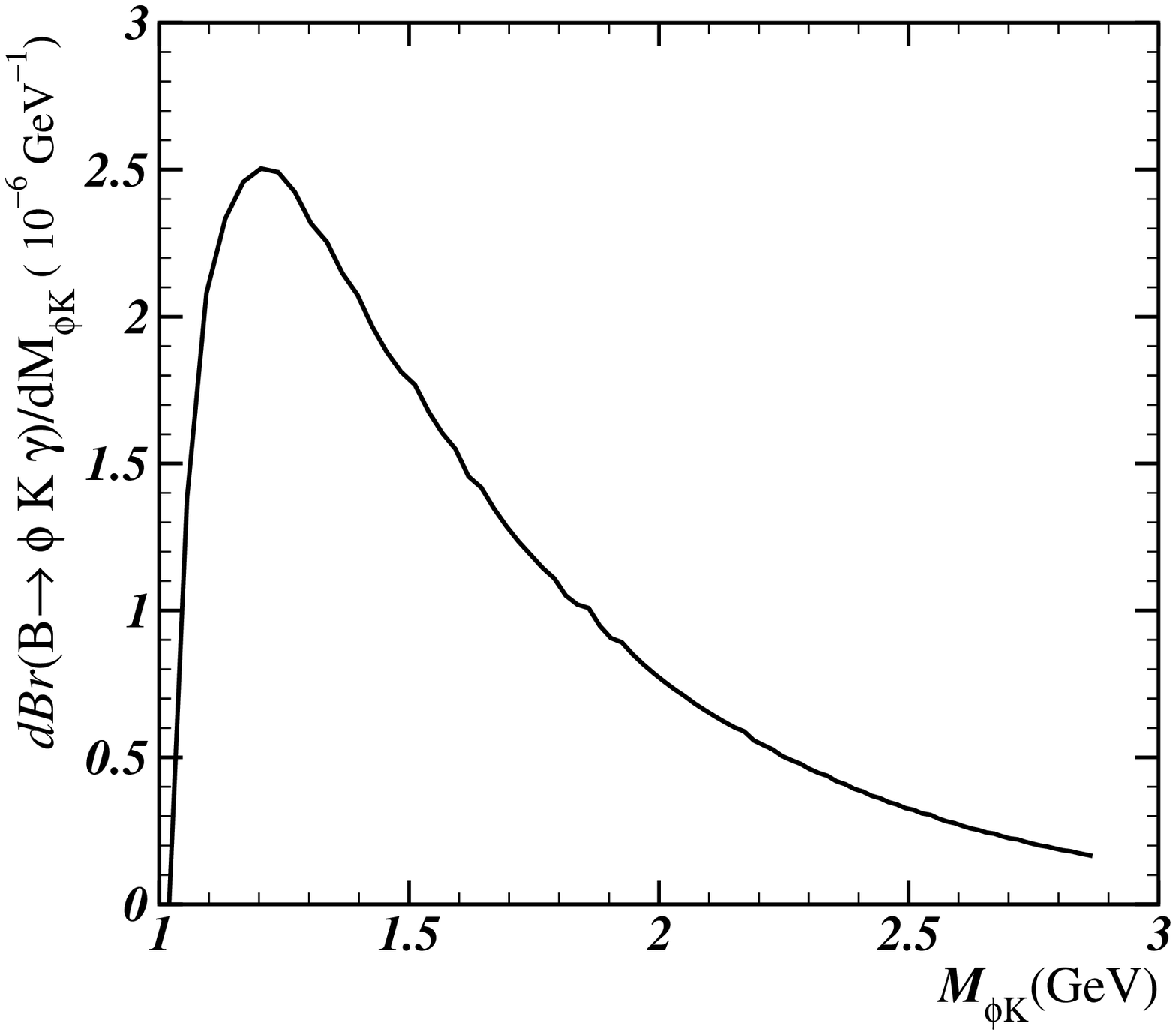}
\includegraphics*[width=3.0
  in]{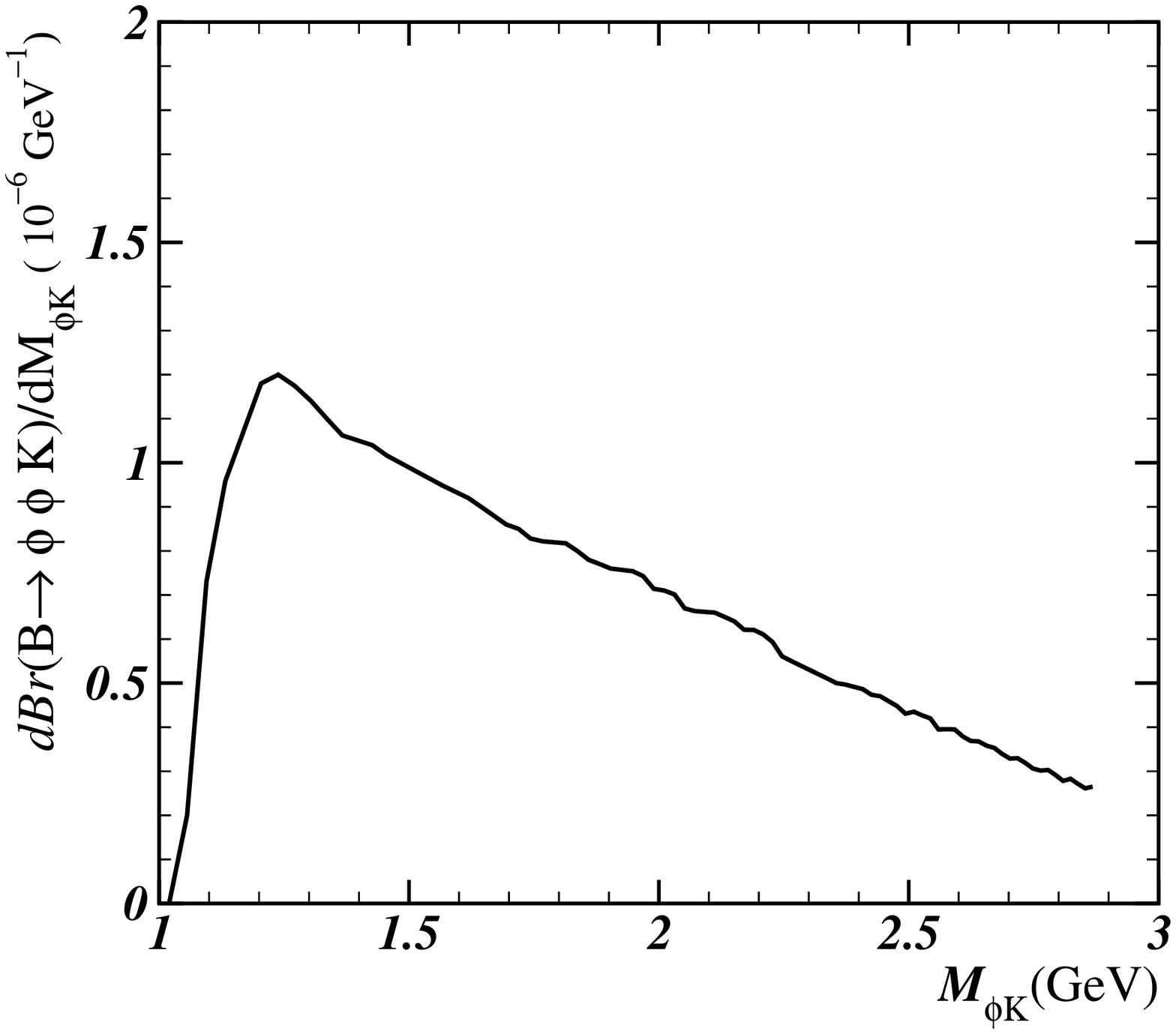}
\caption{$B\to\phi K\gamma$ and $B\to\phi\phi K$ decay spectra
in the $\phi K$ invariant mass.}\label{2phik}
\end{figure}

Because there exists only an upper bound for the measured $\tau
\to \phi K\nu$ branching ratio, we also consider the $\tau \to
K^*\pi\nu$ branching ratio, when constraining the parameter
$m_\parallel$. That is, we assume that $m_\parallel$ in the two
decay modes, i.e., the time-like $\phi K$ and $K^*\pi$ form
factors, do not differ much. The experimental data and the
theoretical prediction from chiral perturbation theory are
\begin{eqnarray}
B(\tau \to \phi K\nu)& < & 6.7\times 10^{-5}\ \ \ \texttt{\cite{PDG}} \;,\nonumber\\
B(\tau \to K^*\pi\nu)& \approx & 2.7\times 10^{-4}\ \ \ \texttt{\cite{Wise}} \;,\nonumber\\
B(B\to \phi K\gamma)&=&(3.4\pm 0.9\pm0.4 )\times 10^{-6}\ \ \
\texttt{\cite{BelA}}\;. \label{da}
\end{eqnarray}
For the application of our formalism to the $\tau$ decays, we shall
trust it up to the order-of-magnitude accuracy, since
the ratio $\eta=w^2/m_\tau^2$ in this case is not a small parameter.
With $m_{\parallel, T}\approx 2$ GeV,
we obtain from Eqs.~(\ref{tau}) and (\ref{gamma}),
\begin{eqnarray}
B(\tau \to \phi K\nu)&=& 6.8\times 10^{-5}\;,\nonumber\\
B(\tau \to K^*\pi\nu)&=& 4.5\times 10^{-4}\;,\nonumber\\
B(B\to \phi K\gamma)&=&(2.9^{+0.7}_{- 0.5})\times 10^{-6}\;,
\label{fit}
\end{eqnarray}
consistent with Eq.~(\ref{da})(up to order of magnitude for the
$\tau$ decay data as stated above). Our results are stable with
respect to the variation of $m_\parallel$ and $m_T$ around few
GeV. Therefore, the theoretical error in Eq.~(\ref{fit}) comes
from the variation of the shape parameter $\omega_B$, which can be
regarded as an estimate of the uncertainty from
hadronic dynamics. The predicted $B\to \phi K\gamma$ decay
spectrum is shown in Fig.~\ref{2phik}, which exhibits a maximum at
the $\phi K$ invariant mass around 1.3 GeV, consistent with our
power counting rules.

After constraining the two-meson distribution amplitudes, we
predict the $B\to\phi\phi K$ decay spectrum in the $\phi K$
invariant mass. For this mode, the amplitude ${\cal M}$ is written
as
\begin{eqnarray}
{\cal M}&=&f_{\phi} V^{*}_{tb}V_{ts}\sum_{i=3}^{5} \Big[{\cal
F}^{P(s)}_{Li} + \epsilon_{T} \cdot \epsilon_{3T}(\phi)
{\cal F}^{P(s)}_{Ti}\Big]\;,  \\
{\cal F}^{P(s)}_{Li}&=& 8\pi C_{F} m^2_{B}
\int^{1}_{0} dx_{1} dz\frac{\Phi_{B}(x_{1})
\Phi_{\parallel}(z,\zeta,w) }{x_{1}zm^{2}_{B}+P_{T}^2}
\nonumber \\
&& \times \Big\{ \Big[
(1+z)\Phi_{\parallel}(z,\zeta,w)+\sqrt{\eta}(1-2z)
\Phi_{p}(z,\zeta,w)
\Big]\frac{\alpha _{s}( t^{(1)}_{e})a_{i}^{(s)}(t^{(1)})}
{zm^2_{B}+P_{T}^2}  \nonumber \\
&& +2\sqrt{\eta}\Phi_{p}(z,\zeta,w) \frac{\alpha _{s}(
t^{(2)})a_{i}^{(s)}(t^{(2)}_{e})}
{x_{1}m^2_{B}}\Big\}\;,\label{fl}\\ 
{\cal F}^{P(s)}_{Ti}&=& 8 r_{\phi} \pi C_{F} m^2_{B} \int^{1}_{0}
dx_{1} dz\frac{\Phi_{B}(x_{1}) }{x_{1}zm^{2}_{B}+P_{T}^2}
\nonumber \\
&& \times \Big\{ \Big[
\Phi_{T}(z,\zeta,w) +\Phi_{3}(z,\zeta,w)\sqrt{\zeta(1-\zeta)}
+z\sqrt{\eta}\Phi_{v}(z,\zeta,w) \Big] \frac{\alpha _{s}(
t^{(1)}_{e})a_{i}^{(s)}(t^{(1)})}
{zm^2_{B}+P_{T}^2}\nonumber \\
&&
- \sqrt{\eta}
\Phi_{v}(z,\zeta,w^2)
\frac{\alpha _{s}( t^{(2)})a_{i}^{(s)}(t^{(2)})}
{x_{1}m^2_{B}}\Big\}\;,\label{ft}
\end{eqnarray}
where $\epsilon_{3T}(\phi)$ denote the polarization vectors of the
$\phi$ meson emitted from the weak vertex. The definitions of the
Wilson coefficients $a_i^{(q)}(t)$ are referred to \cite{CKL1}.
For a similar reason, we have dropped all the $O(\eta)$ terms.
Equations (\ref{gamma}), (\ref{fl}) and (\ref{ft}) represent the
amplitudes of the $B$ meson transition into a $VP$ meson pair
associated with different effective operators. We display
the predicted $B^\pm\to\phi\phi K^\pm$ decay spectrum in
Fig.~\ref{2phik}, which also exhibits a maximum at the $\phi K$
invariant mass around 1.3 GeV. Integrating the spectrum over
$\eta$, we obtain the branching ratio without the resonant
contribution in the $\phi\phi$ channel,
\begin{eqnarray}
B(B^\pm\to\phi\phi K^\pm)=(1.3^{+0.4}_{-0.3})\times 10^{-6}\;.
\end{eqnarray}
The uncertainty arises from the variation of the shape parameter
$\omega_B$ of the $B$ meson distribution amplitude.

We have examined other sources of theoretical uncertainty. The
correction to the branching ratios from the neglected $O(\eta)$
terms is about 10\%. To investigate the uncertainty from different
parametrization of meson distribution amplitudes, we have tried
\begin{eqnarray}
\Phi'_{B}(x)=N'_Bx(1-x)\exp
\left[-\frac{1}{2}\left(\frac{xm_B}{\omega'_B}\right)^2\right]\;.
\label{phib1}
\end{eqnarray}
First, the shape parameter $\omega'_B=0.9$ GeV is determined from the
fit to the value of the $B\to\pi$ transition form factor about
0.3. The model $\Phi'_{B}(x)$ is then employed to fix the
$\phi K$ two-meson distribution amplitudes from the data of the
$B\to \phi K \gamma$ branching ratios. It
is observed that the symmetric $z$ dependence in Eq.~(\ref{2pi}) 
should be modified into
\begin{eqnarray}
z(1-z)\to z(1-z)[1+0.5(1-2z)]\;,\label{newz}
\end{eqnarray}
which is reasonable since the $\phi$ meson is heavier than the
kaon. After going through the above procedure, we predict the
$B\to\phi\phi K$ branching ratio using the distribution amplitudes
in Eqs.~(\ref{phib1}) and (\ref{newz}), and find that the result 
increases only by 8\%. We have also checked the sensitivity of our
prediction to the parametrization of the time-like form factors. 
Obeying the normalization and the asymptotic behavior required by 
PQCD, the models with $(w-m_\phi)^2$ [$(w-m_\phi)^3$] being 
replaced by $w^2-m_\phi^2$ [$(w^2-m_\phi^2)^{3/2}$] are
also allowed. Adopting the $B$ meson distribution amplitude in
Eq.~(\ref{phib}), the $\tau \to \phi K\nu$ and $B\to \phi K \gamma$ 
data just imply a slight increase of the parameters $m_\parallel$ 
and $m_T$ to 3-4 GeV. Then we predict the $B\to\phi\phi K$ branching 
ratio using the new parametrization, which is enhanced only by 12\%. 
The above investigations indicate that the PQCD predictions will be
insensitive to the parametrization of meson distribution
amplitudes, if the procedure of determining meson distribution
amplitudes is followed.

Note that the $B^\pm\to\phi\phi K^\pm$ branching ratio has been
measured to be $B(B^\pm\to\phi\phi K^\pm)=(2.6^{+1.1}_{-0.9}\pm
0.3)\times 10^{-6}$ for a $\phi\phi$ invariant mass below 2.85 GeV
\cite{Bel03}. We suggest that the decay spectrum in the $\phi K$
invariant mass should also be measured (only the spectrum in the
$\phi\phi$ invariant mass was presented in \cite{Bel03}), such
that the dynamics of the $B\to VP$ transition can be explored. To
derive the spectrum in the $\phi\phi$ invariant mass, we need to
define the $VV$ two-meson distribution amplitudes, which will be
discussed in the future.

\vskip 0.5cm

We thank H.Y. Cheng and Y.Y. Charng for useful discussions. This
work was supported in part by the National Science Council of
R.O.C. under Grant No. NSC-92-2112-M-001-003 and No.
NSC-92-2112-M-006-026.


\begin{thebibliography}{99}

\bibitem{Belle} BELLE Coll., A. Garmash {\it et al.}, Phys. Rev. D
{\bf 65}, 092005 (2002).
\bibitem{Bar} BABAR Coll., B. Aubert  {\it et al.}, hep-ex/0206004.
\bibitem{CL02} C.H. Chen and H-n. Li, Phys. Lett. B {\bf 561}, 258
(2003).
\bibitem{MP} D. Muller et al., Fortschr. Physik. {\bf 42}, 101 (1994);
M. Diehl, T. Gousset, B. Pire, and O. Teryaev, Phys. Rev. Lett.
{\bf 81}, 1782 (1998); M.V. Polyakov, Nucl. Phys. {\bf B555}, 231
(1999).
\bibitem{BL} G.P. Lepage and S.J. Brodsky, Phys. Lett. B {\bf 87}, 359
(1979); Phys. Rev. Lett. {\bf 43}, 545 (1979); G.P. Lepage and S.
Brodsky, Phys. Rev. D {\bf 22}, 2157 (1980).
\bibitem{BFL} S.J. Brodsky, Y. Frishman and G.P. Lepage,
and C. Sachrajda, Phys. Lett. B {\bf 91}, 239 (1980).
\bibitem{MR} A.V. Efremov and A.V. Radyushkin, Theor. Math. Phys.
{\bf 42}, 97 (1980); Phys. Lett. B {\bf 94}, 245 (1980);
I.V. Musatov and A.V. Radyushkin, Phys. Rev. D {\bf 56}, 2713 (1997).
\bibitem{DM} A. Duncan and A.H. Mueller, Phys. Lett. B {\bf 90}, 159
(1980); Phys. Rev. D {\bf 21}, 1636 (1980).
\bibitem{CZS} V.L. Chernyak, A.R. Zhitnitsky, and V.G. Serbo,
JETP Lett. {\bf 26}, 594 (1977).
\bibitem{BS} J. Botts and G. Sterman, Nucl. Phys. {\bf B225}, 62 (1989);
H-n. Li and G. Sterman, Nucl. Phys. {\bf B381}, 129 (1992).
\bibitem{NL} H-n. Li, Phys. Rev. D {\bf 64}, 014019 (2001);
M. Nagashima and H-n. Li, hep-ph/0202127; Phys. Rev. D {\bf 67},
034001 (2003).
\bibitem{Keum} Y.Y. Keum, H-n. Li, and A.I. Sanda,
AIP Conf. Proc. {\bf 618}, 229 (2002); Y.Y. Keum and A.I. Sanda,
Phys. Rev. D {\bf 67}, 054009 (2003).
\bibitem{Li03} H-n. Li, Czech. J. Phys. {\bf 53}, 657 (2003); Prog.
Part. Nucl. Phys. {\bf 51}, 85 (2003).
\bibitem{BBNS}  M. Beneke, G. Buchalla, M. Neubert, and C.T. Sachrajda,
Phys. Rev. Lett. {\bf 83}, 1914 (1999);
Nucl. Phys. {\bf B606}, 245 (2001).
\bibitem{YL} C.H. Chang and H-n. Li, Phys. Rev. D {\bf 55}, 5577 (1997);
T.W. Yeh and H-n. Li, Phys. Rev. D {\bf 56}, 1615 (1997).
\bibitem{KLS} Y.Y. Keum, H-n. Li, and A.I. Sanda,
Phys. Lett. B {\bf 504}, 6 (2001); Phys. Rev. D {\bf 63}, 054008 (2001);
Y.Y. Keum and H-n. Li, Phys. Rev. {\bf D63}, 074006 (2001).
\bibitem{LUY} C.D. L\"{u}, K. Ukai, and M.Z. Yang, Phys. Rev. D {\bf 63},
074009 (2001).
\bibitem{CLY} H.Y. Cheng, H-n. Li, and K.C. Yang,
Phys. Rev. D {\bf 60}, 094005 (1999).
\bibitem{LY1} H-n. Li and H.L. Yu, Phys. Rev. Lett. {\bf 74}, 4388 (1995);
Phys. Rev. D {\bf 53}, 2480 (1996).
\bibitem{LU} H-n. Li, Phys. Rev. D {\bf 66}, 094010 (2002);
H-n. Li and K. Ukai, Phys. Lett. B {\bf 555}, 197 (2003).
\bibitem{BSW} M. Bauer, B. Stech, M. Wirbel,
Z. Phys. C {\bf 29}, 637 (1985); Z. Phys. C {\bf 34}, 103
(1987).
\bibitem{CHST}  C.K. Chua, W.S. Hou, and S.Y. Tsai,
Phys. Rev. D{\bf 66}, 054004 (2002); W.S. Hou, S.Y. Shiau, and
S.Y. Tsai, Phys. Rev. D {\bf 67}, 034012 (2003).
\bibitem{BFOPP} B. Bajc, S. Fajfer, R.J. Oakes, T.N. Pham, and
S. Prelov\v sek, Phys. Lett. B {\bf 447}, 313 (1999);
S. Fajfer, T.N. Pham, and A. Prapotnik, hep-ph/0401120.
\bibitem{cheng} H.Y. Cheng and K.C. Yang, Phys. Rev. D {\bf 66},
054015 (2002); Phys. Rev. D {\bf 66}, 094009 (2002).
\bibitem{LGT} O. Leitner, X.H. Guo, and A.W. Thomas,
Eur. Phys. J. C{\bf 31}, 215 (2003).
\bibitem{W} Z.T. Wei, hep-ph/0301174.
\bibitem{Bel03} BELLE Coll., H.C. Huang {\it et al.}, Phys. Rev. Lett.
{\bf 91}, 241802 (2003).
\bibitem{PDG} Particle Data Group, K. Hagiwara {\it et al.}, Phys. Rev.
D {\bf 66}, 010001 (2002).
\bibitem{M} M. Maul, Eur. Phys. J. C {\bf 21}, 115 (2001).
\bibitem{DFK} M. Diehl, Th. Feldmann, P. Kroll, and C. Vogt,
Phys. Rev. D {\bf 61}, 074029 (2000);  M. Diehl, T. Gousset, and
B. Pire, Phys. Rev. D {\bf 62}, 073014 (2000).
\bibitem{PB1} P. Ball, V.M. Braun, Y. Koike, and K. Tanaka,
Nucl. Phys. {\bf B529}, 323 (1998).
\bibitem{PB2} V.M. Braun and I.E. Filyanov, Z Phys. C {\bf 48}, 239
(1990); P. Ball, J. High Energy Phys. {\bf 01}, 010 (1999).
\bibitem{HHZ} T. Huang, X.H. Wu, and M.Z. Zhou, hep-ph/0402100.
\bibitem{TLS} T. Kurimoto, H-n. Li, and A.I.
Sanda, Phys. Rev. D {\bf 65}, 014007 (2002); Phys. Rev. D {\bf
67}, 054028 (2003).
\bibitem{KR} A. Khodjamirian and R. Ruckl, Phys. Rev. D {\bf 58}, 054013
(1998).
\bibitem{PB3} P. Ball, JHEP {\bf 09}, 005 (1998); Z.
Phys. C {\bf 29}, 637 (1985).
\bibitem{GN} A.G. Grozin and M. Neubert, Phys. Rev. D {\bf 55},
272 (1997); M. Beneke and T. Feldmann, Nucl. Phys. {\bf B592}, 3
(2000).
\bibitem{DS} S. Descotes and C.T. Sachrajda, Nucl. Phys. {\bf B625},
239 (2002).
\bibitem{KKQT} H. Kawamura, J. Kodaira, C.F. Qiao, and K. Tanaka,
Phys. Lett. B {\bf 523}, 111 (2001); Erratum-ibid. {\bf 536}, 344
(2002).
\bibitem{LL04} H-n. Li and H.S. Liao, hep-ph/0404050.
\bibitem{Neu03} B.O. Lange and M. Neubert, Phys. Rev. Lett. {\bf 91},
102001 (2003).
\bibitem{BIK} V.M. Braun, D.Yu. Ivanov, G.P. Korchemsky,
Phys. Rev. D {\bf 69}  (2004) 034014.
\bibitem{LMY} Z.T. Wei and M.Z. Yang, Nucl. Phys. {\bf B642}, 263 (2002);
C.D. Lu and M.Z. Yang, Eur. Phys. J. C {\bf 28}, 515 (2003).
\bibitem{BBL}G. Buchalla {\it et al.}, Rev. Mod. Phys. {\bf 68},
1125 (1996).
\bibitem{Wise} H. Davoudiasl and M.B. Wise, Phys. Rev. D{\bf 53},
2523 (1996).
\bibitem{BelA} Belle Collaboration, A. Drutskoy {\it et al.}, Phys. Rev. Lett. {\bf 92}, 051801 (2004).
\bibitem{CKL1} C.H. Chen, Y.Y. Keum, and H-n. Li, Phys. Rev. D {\bf 64},
112002 (2001); Phys. Rev. D {\bf 66}, 054013 (2002).




\end{thebibliography}
\end{document}